\documentclass[sigconf, authorversion]{acmart}
\AtBeginDocument{%
  }

\setcopyright{acmlicensed}
\copyrightyear{2025}
\acmYear{2025}
\acmDOI{XXXXXXX.XXXXXXX}
\acmConference[MMGenSR '25]{International Workshop on Multimodal Generative Search and Recommendation}{November 14, 2025}{Seoul, Korea}
\acmISBN{978-1-4503-XXXX-X/2018/06}

\usepackage{xspace}
\usepackage{subcaption}
\usepackage{multirow}
\usepackage{colortbl}
\usepackage{arydshln}
\usepackage{enumitem}

\providecommand{\inmo}{INMO\xspace}
\providecommand{\neumf}{NeuMF\xspace}
\providecommand{\multvae}{Mult-VAE\xspace}
\providecommand{\ngcf}{NGCF\xspace}
\providecommand{\imcgae}{IMC-GAE\xspace}
\providecommand{\idcf}{IDCF\xspace}
\providecommand{\lgcn}{LightGCN\xspace}
\providecommand{\sgl}{SGL\xspace}
\providecommand{\xsimgcl}{XSimGCL\xspace}
\providecommand{\alpharec}{AlphaRec\xspace}
\providecommand{\unicdr}{UniCDR\xspace}
\providecommand{\ppa}{PPA\xspace}
\providecommand{\ours}{MICRec\xspace}

\newcommand{\set}[1]{\mathcal{#1}}

\providecommand{\sE}{\ensuremath{\set{E}}}
\providecommand{\sI}{\ensuremath{\set{I}}}
\providecommand{\sN}{\ensuremath{\set{N}}}

\providecommand{\sU}{\ensuremath{\set{U}}}

\renewcommand{\vec}[1]{{\bf{#1}}}
\providecommand{\vb}{\ensuremath{\vec{b}}}
\providecommand{\ve}{\ensuremath{\vec{e}}}
\providecommand{\vr}{\ensuremath{\vec{r}}}
\providecommand{\vt}{\ensuremath{\vec{t}}}
\providecommand{\vecv}{\ensuremath{\vec{v}}}
\providecommand{\vx}{\ensuremath{\vec{x}}}

\newcommand{\mat}[1]{\boldsymbol{#1}}
\providecommand{\mW}{\ensuremath{\mat{W}}}

\providecommand{\sIs}{\ensuremath{\set{I}_\text{s}}}
\providecommand{\sIu}{\ensuremath{\set{I}_\text{u}}}
\providecommand{\sItem}{\ensuremath{\set{I}_\text{tem}}}
\providecommand{\sUs}{\ensuremath{\set{U}_\text{s}}}
\providecommand{\sUu}{\ensuremath{\set{U}_\text{u}}}
\providecommand{\sUtem}{\ensuremath{\set{U}_\text{tem}}}

\begin{document}

\title[Unifying Inductive, Cross-Domain, and Multimodal Learning for Robust and Generalizable Recommendation]{Unifying Inductive, Cross-Domain, and Multimodal Learning\\for Robust and Generalizable Recommendation}

\author{Chanyoung Chung}
\email{chanyoung.chung@kaist.ac.kr}
\affiliation{%
  \institution{KAIST}
  \city{Daejeon}
  \country{Republic of Korea}
}
\author{Kyeongryul Lee}
\email{klee0257@kaist.ac.kr}
\affiliation{%
  \institution{KAIST}
  \city{Daejeon}
  \country{Republic of Korea}
}
\author{Sunbin Park}
\email{seonbeen.park@dplanex.com}
\affiliation{%
  \institution{Dplanex}
  \city{Seoul}
  \country{Republic of Korea}
}
\author{Joyce Jiyoung Whang}
\authornote{Corresponding author.}
\email{jjwhang@kaist.ac.kr}
\affiliation{%
  \institution{KAIST}
  \city{Daejeon}
  \country{Republic of Korea}
}


\begin{abstract}
Recommender systems have long been built upon the modeling of interactions between users and items, while recent studies have sought to broaden this paradigm by generalizing to new users and items, incorporating diverse information sources, and transferring knowledge across domains. Nevertheless, these efforts have largely focused on individual aspects, hindering their ability to tackle the complex recommendation scenarios that arise in daily consumptions across diverse domains. In this paper, we present \ours, a unified framework that fuses inductive modeling, multimodal guidance, and cross-domain transfer to capture user contexts and latent preferences in heterogeneous and incomplete real-world data. Moving beyond the inductive backbone of \inmo~\cite{inmo}, our model refines expressive representations through modality-based aggregation and alleviates data sparsity by leveraging overlapping users as anchors across domains, thereby enabling robust and generalizable recommendation. Experiments show that \ours outperforms 12 baselines, with notable gains in domains with limited training data.
\end{abstract}




\maketitle

\section{Introduction}
Recommender systems play a pivotal role in addressing information overload by modeling user–item interactions to deliver personalised experiences~\cite{reasoningrec, unleash}. A widely adopted approach is collaborative filtering with latent factor models, which infer latent user and item preferences and have demonstrated strong performance across diverse applications~\cite{rankformer, hyperhawkes}. However, in domains characterized by multimodal content, dynamic user preferences, and cross-domain dependencies, including but not limited to fashion, food, and lifestyle, the limitations of such approaches become evident~\cite{food,fashion}. These limitations are further compounded in real-world environments by the continual emergence of new users and items, uneven data distributions across domains, and the increasing prevalence of multimodal information~\cite{causaldiffrec,td3, frcsu}.

Recent research has therefore focused on three distinct directions: (i) inductive recommendation, which seeks to generalize to new users and items without retraining~\cite{elfm, ihm}; (ii) cross-domain recommendation, leveraging user behaviors across multiple domains to mitigate limited interactions~\cite{unicdr, disencdr, ppa}; or (iii) multimodal recommendation, which integrates modalities such as text and images to capture semantic aspects of items and users~\cite{modicf, alpharec}. However, existing works have largely focused on a single aspect, either pursuing inductive capability at the expense of modality and domain generalization~\cite{inmo,idcf}, or leveraging multimodal and cross-domain aspects yet falling short in handling unseen\footnote{The terms ``new" and ``unseen" are used interchangeably in the manuscript.} users and items~\cite{motkd, sieoug}, all of which remain insufficient for real-world applications.

To address this problem, we propose \ours, a unified framework integrating \textbf{\underline{M}}ultimodal learning, \textbf{\underline{I}}nductive modeling, and \textbf{\underline{C}}ross-domain transfer into a cohesive model for real-world \textbf{\underline{Rec}}om-mendation scenarios. Extending \inmo~\cite{inmo}, which incorporates template-based initialization, \ours derives more expressive representations by capturing multimodal similarities across users and items, while introducing a contrastive loss on overlapping users to facilitate effective knowledge transfer across domains. Experiments show that \ours outperforms competitive inductive, multimodal, and cross-domain baselines, with average improvements exceeding 14.6\% relative to \inmo. Our main contributions are threefold:

\begin{figure*}[t]
\centering
\includegraphics[width=\linewidth]{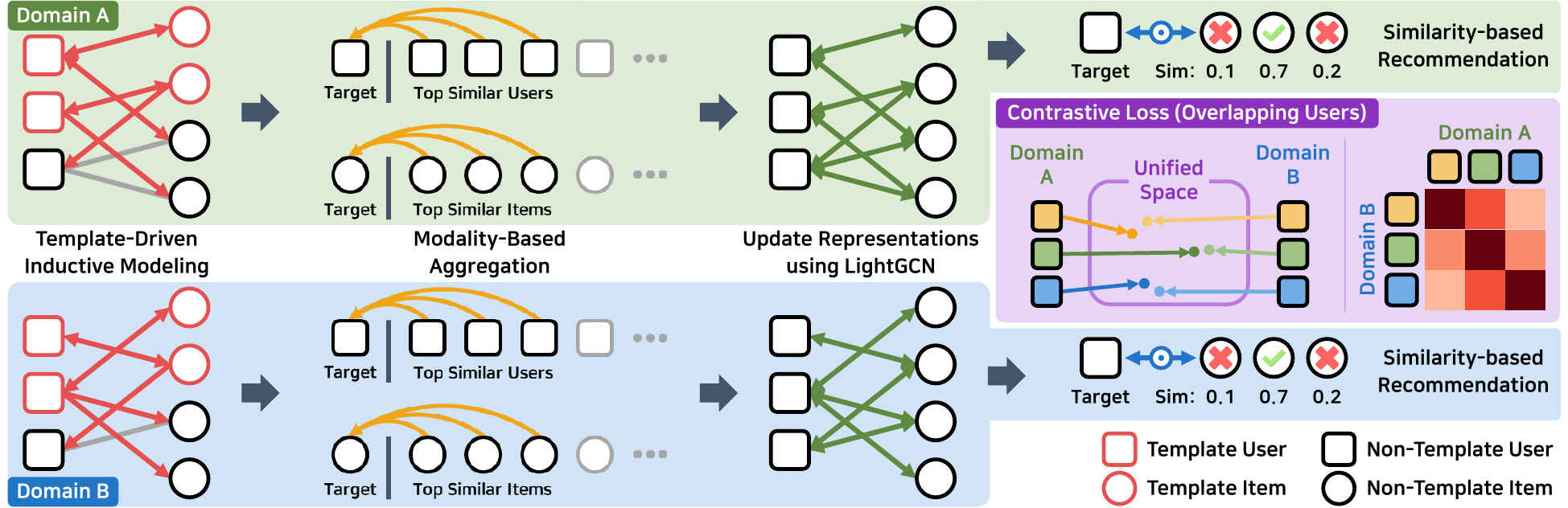}
\caption{Overview of \ours. \ours enables inductive, multimodal, and cross-domain recommendation based on template-driven inductive modeling, modality-based aggregation, and contrastive learning on overlapping users between two domains.}
\label{fig:overview}
\end{figure*}

\begin{itemize}[leftmargin=*]
    \item We present MICRec, a unified framework that bridges inductive, multimodal, and cross-domain paradigms for robust and generalizable recommendation across diverse settings.
    \item We introduce a multimodal similarity-driven representation to strengthen embeddings and incorporate contrastive learning to facilitate cross-domain alignment, thereby enabling the model to overcome limitations in data-scarce domains.
    \item Extensive experiments on our real-world datasets empirically show that MICRec substantially outperforms INMO and other baselines across a wide range of recommendation scenarios.\footnote{We provide our implementations at \url{https://github.com/bdi-lab/MICRec}.}
\end{itemize}

\section{Related Work}
Most recommendation studies have focused on transductive settings, where all users and items are observed during training. Early collaborative filtering models like \neumf~\cite{neumf} and \multvae~\cite{multvae} primarily capture interactions through latent factorization or using variational autoencoder. Subsequent work, such as \ngcf~\cite{ngcf} and \lgcn~\cite{lightgcn}, further extended this direction by exploiting structural relations on user-item graphs, while self-supervised learning approaches~\cite{sgl, xsimgcl} provide additional supervision through contrastive objectives. Alongside these efforts, recent research has broadened the scope with multimodal integration incorporating auxiliary content~\cite{alpharec}, while cross-domain methods~\cite{disencdr, unicdr} transfer knowledge across domains. Despite their success, these methods remain limited by their reliance on transductive assumptions, which restrict generalization to users or items unseen during training.

Faced with the challenge of unseen users and items, inductive recommendation has gained increasing attention, enabling models to adapt without retraining when new entities emerge. To address unseen users, IDCF~\cite{idcf} leverages similarities with key users, while Eflm~\cite{elfm} utilizes user-item co-clustering. ColdGPT~\cite{coldgpt} integrates item attribute graphs with pretrained language models to handle new items, and IHM~\cite{ihm} employs a two-tower structure to alleviate cold-start item challenges. However, these studies typically concentrate on either new users or new items in isolation. In contrast, INMO~\cite{inmo} takes an important step by constructing an inductive framework that simultaneously models both unseen users and unseen items. Building on this foundation, our study introduces a unified framework that further incorporates multimodal guidance and cross-domain transfer to overcome the constraints of neglected diverse modalities and sparse interactions, moving toward a more comprehensive solution for real-world recommendation scenarios.

\section{Problem Statement}
Unlike transductive setting that assumes all users and items to be learned, inductive recommendation allows new entities to appear at inference time. Formally, the user set $\sU$ and item set $\sI$ are partitioned into seen ($\sUs$, $\sIs$) and unseen ($\sUu$, $\sIu$) subsets, respectively. In the training phase, a model is optimized on the observed interactions $\sE_\text{tr}\subseteq\sUs\times\sIs$, while a separate validation set $\sE_\text{val}\subseteq\sUs\times\sIs$ is used for hyperparameter tuning. After training, additional interactions $\sE_\text{new}\subseteq\sU\times\sI$ involving previously unseen users and items are provided, and the model leverages both $\sE_\text{tr}$ and $\sE_\text{new}$ to generate the recommendations on the test interactions $\sE_\text{test}\subseteq\sU\times\sI$.

Building upon this formulation, we extend inductive recommendation to incorporate both multimodal attributes and cross-domain scenarios. Let $\sU^A$ and $\sU^B$ denote the user sets, and $\sI^A$ and $\sI^B$ the item sets, of two domains $A$ and $B$, respectively. In this setting, two domains share a subset of users, denoted by $\sU_\text{o}=\sU^A\cap\sU^B$, while the item sets remain domain-specific. For each domain $D$, each item $i\in\sI^D$ is further associated with multimodal attributes, including a textual description $t_i^D$ and an image $v_i^D$. By integrating multimodal features and collaborative information from both domains, the model aims to recommend relevant items for each user within their respective domain under the inductive scenario.

\section{Methodology}
\inmo~\cite{inmo} represents the most notable inductive approach that simultaneously addresses unseen users and items. To advance this line of research, we extend its framework with multimodal attributes and cross-domain scenarios, and propose a new framework, \ours. Specifically, \ours constructs template-based representations to generalize to unseen users and items (\S~\ref{sec:tem}), refines them through multimodal feature aggregation (\S~\ref{sec:sim}), and employs a contrastive loss on overlapping users to facilitate knowledge transfer across domains (\S~\ref{sec:cl}). An overview of \ours is shown in Figure~\ref{fig:overview}.

\subsection{Template-Driven Inductive Modeling}
\label{sec:tem}
Conventional recommendation models typically rely on learnable embeddings assigned to each user and item. Despite their effectiveness, this design inherently confines them to the transductive setting, as unseen users and items cannot be directly represented. To overcome this limitation, we build on \inmo~\cite{inmo} which introduces template users and items to derive representations for unseen entities. This allows \ours to retain the advantages of learnable embeddings while generalizing effectively to new users and items.

For each domain $D$, we define a set of template users $\sUtem^D\subseteq\sUs^D$ and a set of template items $\sItem^D\subseteq\sIs^D$. The representations of users and items are derived from their connections to the templates as
\begin{equation*}
\vx_u^D = \left(|\sN_u^D\cap\sItem^D|+1\right)^{-\alpha}\sum_{i'\in\sN_u^D\cap\sItem^D}\left(\ve_{i'}^D+\vb_\text{user}^D\right),
\end{equation*}
\begin{equation*}
\vx^D_i = \left(|\sN_i^D\cap\sUtem^D|+1\right)^{-\alpha}\sum_{u'\in\sN_i^D\cap\sUtem^D}\left(\ve_{u'}^D+\vb_\text{item}^D\right),
\end{equation*}
where $\vx_u^D, \vx_i^D\in\mathbb{R}^d$ denote the user and item representations, $d$ is the embedding dimension, $\sN_u^D=\{i:(u,i)\in\sE^D\}$ and $\sN_i^D=\{u:(u,i)\in\sE^D\}$ are the neighbour sets of a user and an item, $\alpha$ is a coefficient controlling the normalization, $\ve_u^D,\ve_i^D\in\mathbb{R}^d$ are learnable embeddings of template users and items, and $\vb_\text{user}^D,\vb_\text{item}^D\in\mathbb{R}^d$ are global biases shared across all users and items, respectively. This formulation, originally proposed in \inmo, enables representations of unseen entities to be derived from their associated templates. We next enhance this inductive backbone with modality-based aggregation and cross-domain knowledge transfer to improve robustness.

\subsection{Modality-Based Aggregation}
\label{sec:sim}
While user-item interaction graphs are effective in modeling structural relationships between users and items, they often fail to capture semantic relevance among users or items. For instance, two semantically similar items may appear far apart in the interaction graph if they are consumed by disjoint user groups. To mitigate this issue, we augment the representations with multimodal feature-based similarities, enabling \ours to capture semantic closeness that is not explicitly revealed by the interaction graph.

For each item $i\in\sI^D$, the textual description $t_i^D$ is encoded with SentenceBERT~\cite{sbert} and the image $v_i^D$ with ViT~\cite{vit}, yielding textual and visual features $\vt_i^D$, $\vecv_i^D$, respectively. Since multimodal attributes are not explicitly provided for users, their multimodal features are inferred from the items with which they have interacted. In particular, the textual feature $\vt_u^D$ and the visual feature $\vecv_u^D$ of a user $u\in\sU^D$ are defined as the averages of the corresponding features of the interacted items. Based on the multimodal features, similarity scores between users and between items are obtained as
\begin{align*}
\text{sim}^D(u,u')&=w\cdot c(\vt_u^D,\vt_{u'}^D)+(1-w)\cdot c(\vecv_u^D,\vecv_{u'}^D), \\
\text{sim}^D(i,i')&=w\cdot c(\vt_i^D,\vt_{i'}^D)+(1-w)\cdot c(\vecv_i^D,\vecv_{i'}^D),
\end{align*}
where $w\in(0,1)$ is a hyperparameter controlling the contributions of textual and visual features, and $c(\cdot,\cdot)$ denotes cosine similarity.

We then enhance the user and item representations based on their most similar counterparts in terms of multimodal similarities. Specifically, for each user $u\in\sU^D$, we define $\widetilde{\sN_u}^D$ as the set of top $K$ most similar users, and for each item $i\in\sI^D$, we define $\widetilde{\sN_i}^D$ as the set of top $K$ most similar items, where $K$ is a hyperparameter that specifies the number of the most similar entities. The representations of users and items are subsequently refined by aggregating those of the most similar users and items, respectively. The aggregation process can be formulated as follows:
\begin{equation*}
\widetilde{\vx_u}^D=\vx_u^D+\frac{1}{K}\sum_{u'\in\widetilde{\sN_u}^D}\vx_{u'}^D,\quad\widetilde{\vx_i}^D=\vx_i^D+\frac{1}{K}\sum_{i'\in\widetilde{\sN_i}^D}\vx_{i'}^D.
\end{equation*}

The resulting representations $\widetilde{\vx_u}^D$ and $\widetilde{\vx_i}^D$ are subsequently fed into \lgcn~\cite{lightgcn}, from which we obtain the final user and item representations $\vr_u^D$ and $\vr_i^D$ for each domain $D$.

\subsection{Cross-Domain Contrastive Learning}
\label{sec:cl}

\noindent\textbf{\textit{BPR Loss}$\ $} When training recommender systems, one of the most widely used loss functions is the Bayesian Personalized Ranking (BPR) loss~\cite{bpr}. The core idea of BPR loss is to ensure that, for each user, interacted items receive higher scores than non-interacted ones. Formally, the score of item $i$ for user $u$ in domain $D$ is defined as $s(u,i)=\vr_u^D\cdot\vr_i^D$, and the BPR loss is formulated as:
\begin{equation*}
\text{\footnotesize$\displaystyle
L_\text{BPR}^D=-\sum_{u\in\sUs^D}\sum_{i\in\sN_u^D}\sum_{i^-\in\sIs^D\setminus\sN_u^D}\ln\sigma(s(\vr_u^D,\vr_i^D)-s(\vr_u^D,\vr_{i^-}^D)) + \lambda\lVert\Theta\rVert_2^2,$%
}
\end{equation*}
where $\Theta$ represents model parameters and $\lambda$ is a regularization coefficient. We also adopt the self-enhanced (SE) loss proposed in \inmo~\cite{inmo} to accelerate the training, which is defined as follows:
\begin{equation*}
\text{\footnotesize$\displaystyle
L_\text{SE}^D=-\sum_{u\in\sUtem^D}\sum_{i\in\sN_u^D\cap\sItem^D}\sum_{i^-\in\sItem^D\setminus\sN_u^D}\ln\sigma(s(\ve_u^D,\mW\ve_i^D)-s(\ve_u^D,\mW\ve_{i^-}^D))$%
}
\end{equation*}
where $\mW\in\mathbb{R}^{d\times d}$ is a projection matrix. The SE loss can be regarded as a variant of the BPR loss that directly employs the embeddings of template users and items for recommendation.

\vspace{0.3em}\noindent\textbf{\textit{Contrastive Loss}$\ $} When the available training data is insufficient, recommender systems become less effective in modeling user-item interactions~\cite{unicdr}. To mitigate this limitation, cross-domain recommender systems leverage information from multiple domains to enable knowledge transfer among them~\cite{disencdr,ppa}. In cross-domain scenarios, overlapping users, who simultaneously exist in both domains, naturally serve as anchors and provide cues for aligning representations across domains. Motivated by this, we design a contrastive loss that brings the representations of the same user across domains closer while pushing apart those of different users.

\begin{table}[t]
\centering
\caption{Statistic of the Generated Datasets.}
\begin{tabular}{ccccccc}
\toprule
 & $|\sU|$ & $|\sI|$ & $|\sE_\text{tr}|$ & $|\sE_\text{new}|$ & $|\sE_\text{val}|$ & $|\sE_\text{test}|$ \\
\midrule
Food & 1,107 & 461 & 10,069 & 4,005 & 2,080 & 3,770 \\
Kitchen & 571 & 491 & 4,190 & 1,948 & 914 & 1,625 \\
\hdashline
Beauty & 973 & 645 & 9,588 & 3,475 & 1,909 & 3,515 \\
Electronics & 9,729 & 6,065 & 74,597 & 33,075 & 15,710 & 28,873 \\
\hdashline
Toy & 424 & 433 & 3,102 & 1,587 & 1,017 & 1,248 \\
Game & 832 & 709 & 7,002 & 2,719 & 2,003 & 2,599 \\
\bottomrule
\end{tabular}
\label{tab:stat}
\end{table}

\begin{table*}[t!]
\small
\centering
\setlength{\tabcolsep}{0.3em}
\caption{Performance Comparison Between \ours and Baseline Methods. The best results are shown in bold, while the second-best are underlined. Improv. (\%) indicates the relative performance improvement of \ours with respect to \inmo.}
\vspace{-0.5em}
\begin{tabular}{ccccccccccccccccccc}
\hline
 & \multicolumn{6}{c}{\textit{Food\&Kitchen}} & \multicolumn{6}{c}{\textit{Beauty\&Electronics}} & \multicolumn{6}{c}{\textit{Toy\&Game}} \\
 & \multicolumn{3}{c}{Food} & \multicolumn{3}{c}{Kitchen} & \multicolumn{3}{c}{Beauty} & \multicolumn{3}{c}{Electronics} & \multicolumn{3}{c}{Toy} & \multicolumn{3}{c}{Game} \\
 & Pre & Rec & NDCG & Pre & Rec & NDCG & Pre & Rec & NDCG & Pre & Rec & NDCG & Pre & Rec & NDCG & Pre & Rec & NDCG \\
\hline
\neumf & 2.660 & 16.218 & 9.367 & 1.436 & 11.161 & 6.790 & 3.566 & 16.700 & 11.320 & 0.671 & 4.705 & 2.633 & 1.226 & 7.838 & 3.727 & 1.983 & 12.613 & 7.285 \\
\multvae & \underline{4.232} & \underline{24.355} & 14.175 & 2.513 & 17.065 & 10.948 & 4.723 & 23.395 & 16.304 & 1.155 & 7.781 & 4.606 & 1.934 & 12.303 & 6.764 & 3.630 & 23.786 & 14.067 \\
\ngcf & 2.977 & 17.192 & 9.843 & 2.636 & 16.211 & 7.922 & 5.283 & 26.182 & 16.439 & 0.637 & 4.269 & 2.355 & 2.134 & 13.726 & 7.547 & 3.101 & 20.275 & 11.260 \\
\imcgae & 3.686 & 21.577 & 12.453 & 1.821 & 12.939 & 7.548 & 4.450 & 21.894 & 14.881 & 0.886 & 5.954 & 3.459 & 1.262 & 8.157 & 4.082 & 3.119 & 20.018 & 11.434 \\
\idcf & 3.460 & 20.357 & 11.584 & 3.345 & 21.937 & 13.953 & 5.396 & 27.852 & 17.502 & 1.052 & 6.877 & 4.044 & 2.087 & 12.948 & 7.029 & 3.341 & 21.843 & 11.698 \\
\lgcn & 4.142 & 24.263 & \underline{14.380} & 2.426 & 16.770 & 10.565 & 4.651 & 23.698 & 15.052 & 1.180 & 7.833 & 4.513 & 2.123 & 13.184 & 7.653 & 3.648 & 23.518 & 13.131 \\
\unicdr & 0.709 & 3.841 & 1.956 & 0.665 & 4.436 & 2.055 & 0.457 & 2.410 & 1.291 & 0.048 & 0.255 & 0.125 & 2.238 & 9.366 & 6.034 & \textbf{4.854} & 19.867 & 13.894 \\
\ppa & 1.025 & 5.349 & 2.669 & 0.902 & 5.815 & 2.659 & 1.213 & 6.481 & 3.284 & 0.117 & 0.678 & 0.428 & \underline{2.561} & 12.263 & 7.748 & \underline{4.361} & 18.093 & 12.097 \\
\sgl & 1.129 & 6.115 & 3.132 & 0.928 & 6.157 & 3.010 & 2.169 & 10.992 & 6.023 & 0.185 & 1.216 & 0.614 & 0.955 & 5.982 & 2.952 & 1.400 & 8.812 & 4.844 \\
\xsimgcl & 2.200 & 11.831 & 6.718 & 2.356 & 14.742 & 8.504 & 5.411 & 26.889 & 18.079 & 0.429 & 2.836 & 1.594 & 1.592 & 10.567 & 5.577 & 2.806 & 18.472 & 10.524 \\
\alpharec & 4.038 & 23.534 & 13.865 & \underline{3.608} & \underline{23.491} & 13.613 & \underline{6.552} & \underline{33.876} & \underline{21.625} & 1.253 & 8.543 & 4.853 & 2.512 & \underline{15.770} & \underline{8.503} & 3.972 & \underline{26.610} & \underline{14.852} \\
\hdashline
\inmo & 3.848 & 22.611 & 13.240 & 3.468 & 22.926 & \underline{14.596} & 6.192 & 32.092 & 20.425 & \underline{1.359} & \underline{9.091} & \underline{5.398} & 1.981 & 11.886 & 7.165 & 3.846 & 25.420 & 14.017 \\
\ours & \textbf{4.453} & \textbf{25.773} & \textbf{15.036} & \textbf{3.687} & \textbf{24.289} & \textbf{15.148} & \textbf{6.793} & \textbf{34.777} & \textbf{22.996} & \textbf{1.428} & \textbf{9.591} & \textbf{5.727} & \textbf{2.653} & \textbf{17.202} & \textbf{10.093} & 4.291 & \textbf{28.353} & \textbf{15.889} \\
\rowcolor{gray!20}Improv. (\%) & 15.7 & 14.0 & 13.6 & 6.3 & 5.9 & 3.8 & 9.7 & 8.4 & 12.6 & 5.1 & 5.5 & 6.1 & 33.9 & 44.7 & 40.9 & 11.6 & 11.5 & 13.4 \\
\hline
\end{tabular}
\label{tab:main}
\end{table*}

Formally, given a set $\sU_\text{o}$ of the overlapping users, the loss for matching the corresponding user representations in domain $B$ based on the representations in domain $A$ is defined as follows:
\begin{equation*}
L_\text{1}=-\frac{1}{|\sU_\text{o}|}\sum_{u\in\sU_\text{o}}\log\frac{\exp(f^A(\vr_u^A)\cdot f^B(\vr_u^B)/\tau)}{\sum_{u'\in\sU_\text{o}\setminus\{u\}}\exp(f^A(\vr_u^A)\cdot f^B(\vr_{u'}^B)/\tau)},
\end{equation*}
where $\tau$ is a hyperparameter controlling the temperature. Since the representations of each domain are learned independently, we project the obtained representations into the unified space using $f^A$ and $f^B$, both implemented with 2-layer feedforward network. Similarly, the matching loss from $B$ to $A$ is defined as follows:
\begin{equation*}
L_\text{2}=-\frac{1}{|\sU_\text{o}|}\sum_{u\in\sU_\text{o}}\log\frac{\exp(f^A(\vr_u^A)\cdot f^B(\vr_u^B)/\tau)}{\sum_{u'\in\sU_\text{o}\setminus\{u\}}\exp(f^A(\vr_{u'}^A)\cdot f^B(\vr_u^B)/\tau)}
\end{equation*}
The overall contrastive loss is computed by $L_\text{CL}=L_\text{1}+L_\text{2}$.

\vspace{0.1em}\noindent\textbf{\textit{Joint Loss}$\ $} The overall training objective of \ours is defined as $L=(L_\text{BPR}^A + L_\text{BPR}^B) + \beta(L_\text{SE}^A+L_\text{SE}^B)+\gamma L_\text{CL}$, where $\beta$ and $\gamma$ are hyperparameters balancing SE loss and CL loss, respectively.

\section{Experiments}
We evaluate \ours against state-of-the-art recommendation models on three datasets specifically designed to capture inductive, multimodal, and cross-domain scenarios simultaneously.

\subsection{Experimental Details}

\noindent\textbf{\textit{Datasets}$\ $} For all experiments, we construct three cross-domain datasets based on Amazon Reviews~\cite{amazon}\footnote{\url{https://cseweb.ucsd.edu/~jmcauley/datasets/amazon/links.html}}: \textit{Food(Grocery and Gourmet Food)\&Kitchen(Home and Kitchen)}, \textit{Beauty(Beauty)\&Electronics(Elec-tronics)}, and \textit{Toy(Toys and Games)\&Game(Video Games)}. Following prior works on recommender systems~\cite{inmo,motkd}, we retain only interactions with ratings of 4 or higher, and keep users and items that are involved in more than 10 interactions.

To simulate the inductive scenario, we follow the data processing of \inmo~\cite{inmo}. For each domain, 20\% of users and items are randomly sampled as unseen, and all associated interactions are removed from the training set. The removed interactions are regarded as $\sE_\text{new}$, and introduced as additional interactions at inference time.

\vspace{0.5em}\noindent\textbf{\textit{Baseline Methods}$\ $} We compare \ours against 12 baselines, including traditional collaborative filtering methods as well as multimodal, cross-domain, and inductive recommendation models. The baselines include \neumf~\cite{neumf}, \multvae~\cite{multvae}, \ngcf~\cite{ngcf}, \imcgae~\cite{imcgae}, \idcf~\cite{idcf}, \lgcn~\cite{lightgcn}, \unicdr~\cite{unicdr}, \ppa~\cite{ppa}, \sgl~\cite{sgl}, \xsimgcl~\cite{xsimgcl}, \alpharec~\cite{alpharec}, and \inmo~\cite{inmo}.

For single-domain methods, we train them independently on each domain and report the performance accordingly. For transductive models, we provide embeddings for unseen users and items by following the initialization strategy specified by each method.

\vspace{0.5em}\noindent\textbf{\textit{Implementation Details}$\ $} We implement \ours in PyTorch~\cite{torch}. The embedding dimension is set to 64, and the number of \lgcn layers is fixed to 3. In all experiments, we set $K=3$, $w=0.9$, $\beta=0.01$, and $\gamma=1.0$. In addition, $\tau$ is set to 0.2 for \textit{Food\&Kitchen}, and 0.1 for \textit{Beauty\&Electronics} and \textit{Toy\&Game}. We use MiniLM-6L~\cite{text} and ViT~\cite{vit} pretrained on ImageNet-21k~\cite{inet} as textual and visual encoders, respectively. Following \inmo~\cite{inmo}, the normalization coefficient $\alpha$ is gradually increased from 0.5 to 1.0, and dropout is applied to prevent overfitting. \ours is optimized using Adam optimizer for up to 1,000 epochs with early stop patience of 50. The number of batches is set to 100, and we randomly sampled 64 overlapping users for each batch to compute the contrastive loss.

\subsection{Quantitative Performance Comparison}
We evaluate each method using three widely adopted metrics: Precision(Pre@$N$), Recall(Rec@$N$), and NDCG(NDCG@$N$). We set $N$ to 20, which means each method recommends 20 items per user, and report all metric values multiplied by 100 for clarity.

\begin{table}[t]
\centering
\setlength{\tabcolsep}{0.52em}
\caption{Ablation Studies of \ours on \textit{Food\&Kitchen}. The best results are shown in bold. Overall, \ours outperforms its variants, highlighting the effectiveness of each module.}
\begin{tabular}{ccccccc}
\toprule
 & \multicolumn{3}{c}{Food} & \multicolumn{3}{c}{Kitchen} \\
 & Pre & Rec & NDCG & Pre & Rec & NDCG \\
\midrule
\inmo & 3.848 & 22.611 & 13.240 & 3.468 & 22.926 & 14.596 \\
w/o MM & 4.146 & 24.228 & 14.167 & 3.555 & 23.313 & 14.768 \\
w/o CD & 4.363 & 25.039 & 14.759 & 3.485 & 22.968 & 14.636 \\
w/o VIS & 4.350 & 25.201 & 14.900 & 3.651 & 23.994 & \textbf{15.218} \\
w/o TXT & 4.232 & 24.713 & 14.529 & 3.651 & 24.141 & 15.205 \\
\hdashline
\ours & \textbf{4.453} & \textbf{25.773} & \textbf{15.036} & \textbf{3.687} & \textbf{24.289} & 15.148 \\
\bottomrule
\end{tabular}
\label{tab:ab}
\end{table}

\begin{table}[t]
\centering
\caption{Recommendation Performance on Low-Degree Items. Results are reported on the bottom 25\% of items ranked by frequency, and \ours consistently outperforms \inmo.}
\begin{tabular}{ccccc}
\toprule
 & \multicolumn{2}{c}{Food} & \multicolumn{2}{c}{Kitchen} \\
 & \inmo & \ours & \inmo & \ours \\
\midrule
Pre@20 & 0.000 & \textbf{0.163} & 0.061 & \textbf{0.123} \\
Rec@20 & 0.000 & \textbf{2.334} & 1.230 & \textbf{2.186} \\
NDCG@20 & 0.000 & \textbf{0.818} & 0.311 & \textbf{0.588} \\
\midrule
 & \multicolumn{2}{c}{Beauty} & \multicolumn{2}{c}{Electronics} \\
 & \inmo & \ours & \inmo & \ours \\
\midrule
Pre@20 & 0.745 & \textbf{0.817} & 0.002 &\textbf{0.015} \\
Rec@20 & 7.827 & \textbf{9.379} & 0.016 & \textbf{0.160} \\
NDCG@20 & 2.367 & \textbf{3.069} & 0.007 & \textbf{0.062} \\
\midrule
 & \multicolumn{2}{c}{Toy} & \multicolumn{2}{c}{Game} \\
 & \inmo & \ours & \inmo & \ours \\
\midrule
Pre@20 & 0.000 & \textbf{0.279} & 0.292 &\textbf{0.556} \\
Rec@20 & 0.000 & \textbf{4.292} & 3.912 & \textbf{7.796} \\
NDCG@20 & 0.000 & \textbf{1.317} & 1.187 & \textbf{2.747} \\
\bottomrule
\end{tabular}
\label{tab:low}
\end{table}

\begin{figure}[t]
\centering
\begin{subfigure}{0.48\columnwidth}
\includegraphics[width=\linewidth]{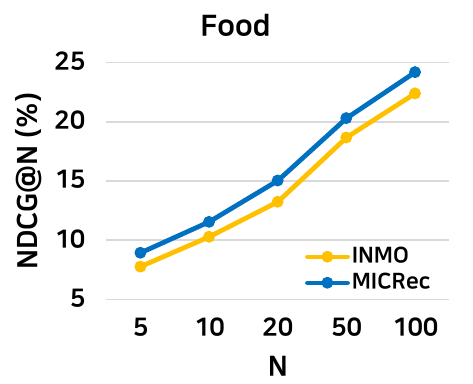}
\caption{Food}
\label{fig:food}
\end{subfigure}
\begin{subfigure}{0.48\columnwidth}
\includegraphics[width=\linewidth]{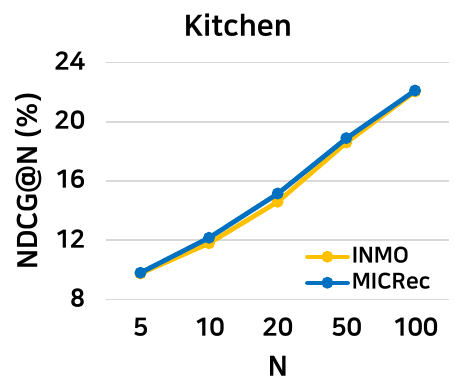}
\caption{Kitchen}
\label{fig:kitchen}
\end{subfigure}
\begin{subfigure}{0.48\columnwidth}
\includegraphics[width=\linewidth]{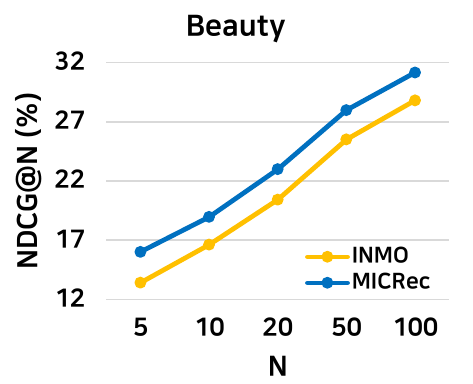}
\caption{Beauty}
\label{fig:beauty}
\end{subfigure}
\begin{subfigure}{0.48\columnwidth}
\includegraphics[width=\linewidth]{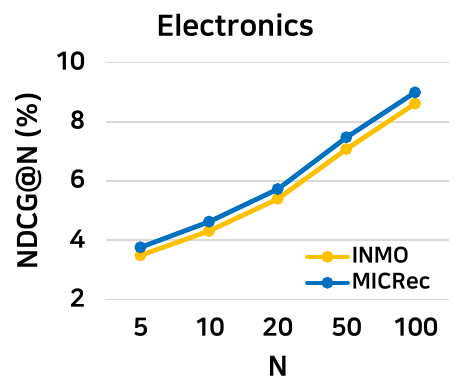}
\caption{Electronics}
\label{fig:electronics}
\end{subfigure}
\begin{subfigure}{0.48\columnwidth}
\includegraphics[width=\linewidth]{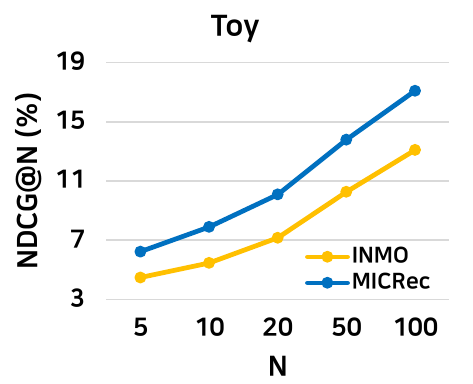}
\caption{Toy}
\label{fig:toy}
\end{subfigure}
\begin{subfigure}{0.48\columnwidth}
\includegraphics[width=\linewidth]{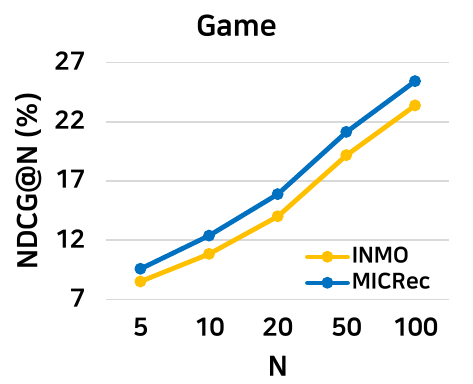}
\caption{Game}
\label{fig:game}
\end{subfigure}
\caption{Comparison of \ours and \inmo on NDCG@$N$ across Varying Number of Recommended Items ($N$). \ours consistently outperforms \inmo across all domains.}
\label{fig:ndcg}
\end{figure}

\begin{figure*}[t]
\centering
\includegraphics[width=\linewidth]{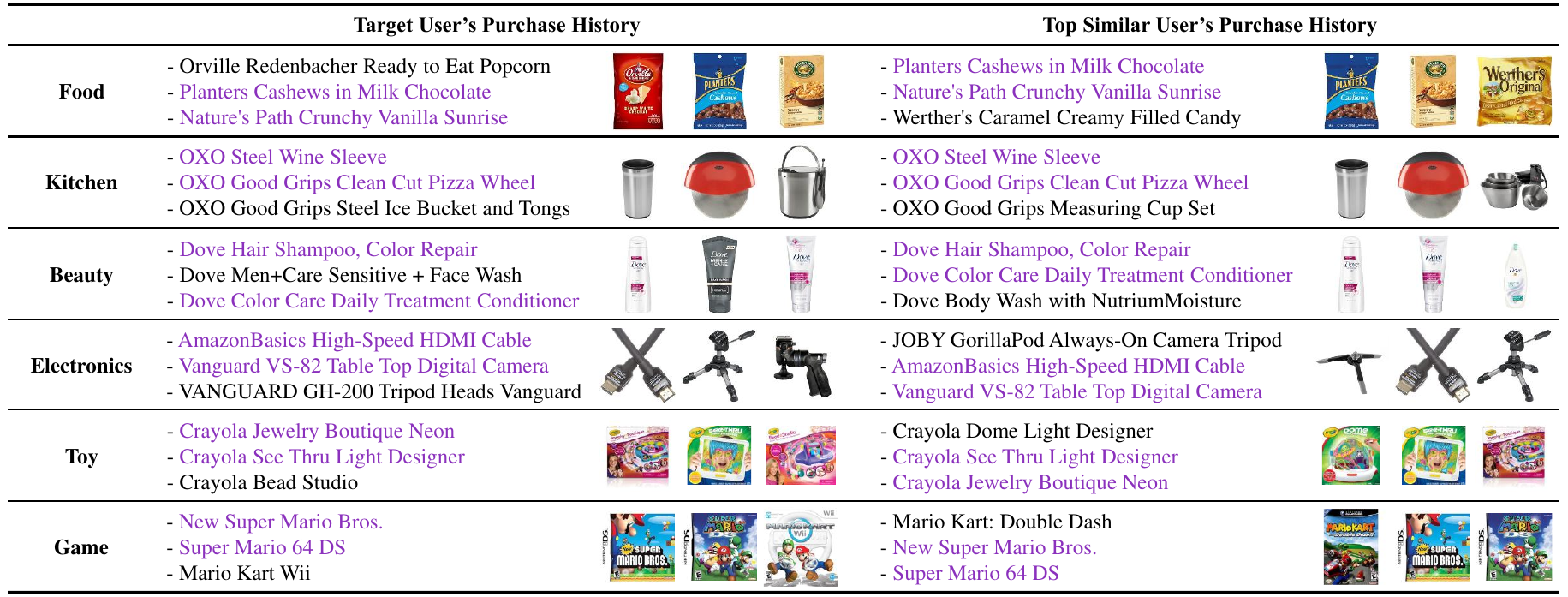}
\caption{Case Study of User Similarity. The images correspond to the products listed in the purchase histories. In all domains, target users and their most similar users exhibit consistent purchase patterns, with overlapping items highlighted in purple.}
\label{fig:case}
\end{figure*}

As shown in Table~\ref{tab:main}, \ours consistently outperforms all baseline methods across all domains and metrics. Moreover, the inductive models, \ours and \inmo, surpass the transductive baselines, highlighting the limitations of existing approaches in handling unseen users and items. The only exception is \alpharec, which employs LLM-based encodings rather than learnable embeddings, enabling it to construct representations for unseen items by exploiting textual features, though its effectiveness remains limited. Moreover, we further evaluate its relative improvement over \inmo. The results demonstrate that \ours achieves an average performance gain of 14.6\%, thereby confirming the effectiveness of the proposed modality-based aggregation and contrastive loss.

\subsection{Effectiveness of \ours over \inmo}
To further investigate the contribution of each module in \ours, we evaluate the recommendation performance of the following variants of \ours on \textit{Food\&Kitchen}: (1) remove multimodal-based aggregation (w/o MM) (2) remove cross-domain contrastive learning (w/o CD) (3) use only textual features for recommendation (w/o VIS) (4) use only visual features for recommendation (w/o TXT). In Table~\ref{tab:ab}, \ours consistently achieves better performance than all of its variants, demonstrating the effectiveness of integrating the multimodal-based aggregation and cross-domain contrastive learning modules. Moreover, all variants exceed the performance of \inmo, further verifying that each component in \ours contributes substantially to the overall recommendation quality.

Low-degree items with limited interactions are inherently susceptible to data sparsity, which constrains the expressiveness of user and item representations. This challenge can be effectively mitigated through multimodal learning and cross-domain knowledge transfer. As shown in Table~\ref{tab:low}, MICRec consistently outperforms INMO on the bottom 25\% of items ranked by frequency. This improvement demonstrates that MICRec alleviates sparsity by aggregating representations from semantically related neighbors through multimodal similarities and by promoting cross-domain transfer with a contrastive loss applied to overlapping users. 

Furthermore, to assess the robustness of MICRec, we varied the number of recommended items $N$ from 5 to 100 and evaluated performance using NDCG@$N$. Figure~\ref{fig:ndcg} illustrates that MICRec consistently surpasses INMO across all values of $N$, further confirming that the integration of multimodal similarity-based representation learning and cross-domain transfer yields robust improvements in overall recommendation performance.

\subsection{Qualitative Study of Multimodal Similarity}
In this section, we conduct a qualitative analysis to demonstrate the effectiveness of similarity-based modality-level aggregation within the multimodal learning of MICRec. In Section~\ref{sec:sim}, as explicit multimodal attributes are unavailable for users, we derive user features as the average of the multimodal features of the items they have interacted with. These features are then used to compute similarity scores between users. To examine whether such similarity reflects semantic relevance between users, we compared the purchase history of each target user with that of its most similar counterpart in each domain. As illustrated in Figure~\ref{fig:case}, target users and their top similar users exhibit highly consistent purchase patterns, including overlapping items, demonstrating that the proposed similarity measure effectively captures semantic relevance between users.

\section{Conclusion and Future Work}
In this work, we proposed MICRec, a unified framework that integrates inductive, multimodal, and cross-domain paradigms to deliver generalizable recommendation across a wide range of real-world domains. MICRec extends INMO by exploiting its inductive capability and incorporating multimodal features to capture fine-grained user and item preferences, while leveraging cross-domain knowledge transfer to address the limitations of recommendation in data-scarce settings. Our empirical results demonstrate the strong generalization capability of MICRec and validate its robustness in low-data regimes. As a next step, we plan to further extend MICRec by leveraging large language models and large multimodal models to ensure reliable performance even under extreme conditions where metadata availability is highly constrained. In addition, we aim to incorporate external knowledge sources, such as knowledge graphs~\cite{vista, hynt, maypl, reed, ingram, smpnn}, to further enhance both representation learning and reasoning process in recommendation.

\begin{acks}
This work was supported by the National Research Foundation of Korea(NRF) grant funded by the Korea government(MSIT)(RS-2025-00559066, Responsible Multimodal Graph AI).
\end{acks}

\bibliographystyle{ACM-Reference-Format}
\bibliography{micrec_mmgensr_2025}


\end{document}